\documentclass[a4paper, 10pt]{jpconf}

\usepackage{relsize}		
\usepackage{url}			
\usepackage{epstopdf}	
\usepackage{graphicx}
\usepackage{tikz}
\usetikzlibrary{decorations.fractals}
\usepackage{enumitem}
\usepackage{pifont}
\usepackage{youngtab}
\usepackage{amssymb}
\usepackage{amsmath}
\usepackage[all]{xy}
\usepackage{amsthm}
\usepackage{verbatim}
\usepackage{graphics}
\usepackage{chemmacros}
\usepackage{shadowtext}
\usepackage{bibentry}
\usepackage{tgbonum}
\usepackage{float}
\usepackage{upgreek}
\usepackage{float}

\definecolor{brinkpink}{rgb}{0.98, 0.38, 0.5}
\definecolor{bluegreen}{RGB}{3, 166, 155}
\definecolor{pitchblack}{RGB}{0, 0, 0}
\definecolor{lightbeige}{RGB}{255, 251, 241}
\definecolor{mediumgray}{RGB}{183, 183, 183}
\definecolor{cerulean}{rgb}{0.16, 0.32, 0.75}

\begin{document}

\title{Characterization of passive CMOS sensors with RD53A pixel modules
}
\author{
\small{Franz Glessgen$^1$, Malte Backhaus$^1$, Florencia Canelli$^2$, Yannick Manuel Dieter$^3$, Jochen Christian Dingfelder$^3$, Tomasz Hemperek$^3$, Fabian Huegging$^3$, Arash Jofrehei$^2$, Weijie Jin$^2$, Ben Kilminster$^2$, Anna Macchiolo$^2$, Daniel Muenstermann$^4$, David-Leon Pohl$^3$, Branislav Ristic$^1$, Rainer Wallny$^1$, Tianyang Wang$^3$, Norbert Wermes$^3$, Pascal Wolf$^3$}}
\address{$^1$ Institute for Particle Physics and Astrophysics, ETH Z\"urich, Z\"urich, Switzerland}
\address{$^2$ Universit\"at Z\"urich, Z\"urich, Switzerland}
\address{$^3$ University of Bonn, Physikalisches Institut, Bonn, Germany}
\address{$^4$ RheinMain University of Applied Sciences, Wiesbaden, Germany}
\ead{franzg@ethz.ch}
\begin{abstract}
Both the current upgrades to accelerator-based HEP detectors (e.g. ATLAS, CMS) and also future projects (e.g. CEPC, FCC) feature large-area silicon-based tracking detectors. We are investigating the feasibility of using CMOS foundries to fabricate silicon radiation detectors, both for pixels and for large-area strip sensors. 
A successful proof of concept would open the market potential of CMOS foundries to the HEP community, which would be most beneficial in terms of availability, throughput and cost.
In addition, the availability of multi-layer routing of signals will provide the freedom to optimize the sensor geometry and the performance, with biasing structures implemented in poly-silicon layers and MIM-capacitors allowing for AC coupling.
A prototyping production of strip test structures and RD53A compatible pixel sensors was recently completed at LFoundry in a 150nm CMOS process.
This presentation will focus on the characterization of pixel modules, studying the performance in terms of charge collection, position resolution and hit efficiency with measurements performed in the laboratory and with beam tests. We will report on the investigation of RD53A modules with 25x100 $\upmu$m$^2$ cell geometry.
\end{abstract}

\section{Introduction}

The High-Luminosity upgrade of the Large Hadron Collider (HL-LHC) will bring the instantaneous luminosity of the LHC to $\mathcal{L} = 5 - 7.5 \times 10^{34}$ cm$^{-2}$ s$^{-1}$ and the pileup to between 140 and 200 for a target integrated luminosity of 3 to 4 ab$^{-1}$ over more than 10 years of operation.
The CMS tracker will have to endure a fluence of $2 \times 10^{16}$ MeV n$_{\text{eq}}$.cm$^{-2}$ for its first layer (at a radius of $r \approx 3$ cm from the interaction region) and a TID of $12$ MGy \cite{TID_info}.
To cope with the expected irradiation levels and the increased trigger rate, the CMS Inner Tracker (IT)  will go through a complete replacement, known as the Phase II upgrade. 
Pixel sensor prototypes were developed in collaboration with LFoundry  \cite{LF} to address the challenges of operating in the Phase II Tracker system.
The samples are passive planar n-in-p sensors for hybrid detector modules built in CMOS technology using the 150 nm production line of LFoundry. 
CMOS technology is largely used in the semiconductor industry and large-scale production of sensors could significantly lower the cost and increase the production speed compared to current standard processes utilized in HEP. 
The considered samples have a 150 $\upmu$m thickness with a pixel size of 25$\times$100 $\upmu$m$^2$ and are DC coupled to the bump bonds of the RD53A readout chip (ROC). The RD53A ROC is a prototype for the developement of the final readout chip for the Phase II upgrade of the IT \cite{RD53_manual}.
The measurements were partly done using the DESY testbeam (TB) infrastructures \cite{DESY}.

\section{CMOS sensors and stitching}
\label{CMOSstitching}
The wafers produced by LFoundry are produced using stitching, a technology that allows the designer to fabricate an image sensor that is larger than the field of view of the lithographic equipment. This is achieved by implementing different building blocks of the design in a reticule cell. By a dedicated  programming of the lithographic tool, each individual block of the reticle can be selected  and can be transferred into the photo-resist on the wafer, possibly with different orientations.  It is then possible to ``stitch'' the various blocks together on the wafer during the lithographic process.
In this way, sensor wafers of several square cm of area, as normally implemented in HEP experiments, can be built.
Additionally, CMOS technology allows for the implementation of very small on-pixel structures which gives access to new sensor features (for HEP). One of them is the addition of metal layers on top of the sensor for signal redistribution. Further, AC-coupled sensors can be produced which cancels the need for leakage current compensation circuits in the preamplifier of the ROC. Also, low and high resistivity polysilicon layers can be deposited on the sensor to be used as field plates and bias resistors respectively.

\section{Qualification measurements}

The sensors used in the IT upgrade have to match performance criteria. Those are summarized in Table \ref{Phase2req}. The following measurements were performed to compare the performance of the LFoundry sensors with the needed requirements.
 All measurements and requirements are quoted for a sensor at a temperature of 20 $^o$C.

\begin{table}[H]
\begin{minipage}{\textwidth}
\caption{\label{Phase2req}Phase II requirements for IT planar sensors. V$_{\text{dep}}$ denotes the full depletion voltage. }
\begin{center}
\begin{tabular}{ll}
\br
Parameter &Requirement\\
\mr
Breakdown voltage & $>$ 350 V \\
Leakage current & $<$ 0.75 $\upmu$A.cm$^{-2}$ at V$_\text{dep}$ + 50 V\\
Efficiency & $>$ 99 \% at V$_\text{dep}$ + 50 V\\
\br
\end{tabular}
\end{center}
\end{minipage}
\end{table}

\subsection{IV}
\label{Sec_IV}

\begin{figure}[H]
\begin{minipage}{\textwidth}
\begin{center}
\includegraphics[scale = 0.19]{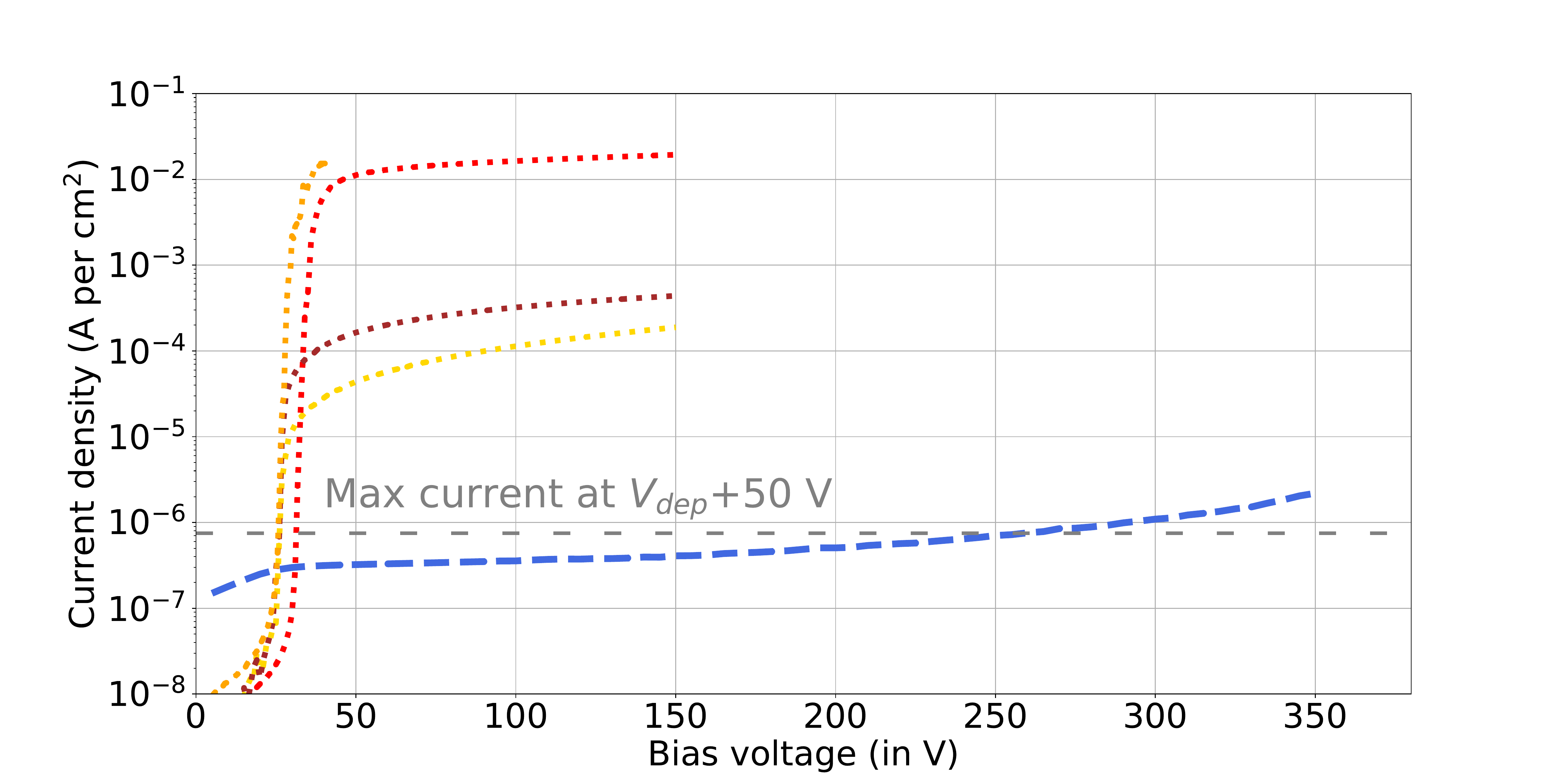}
\end{center}
\caption{\label{IV}IV measurements of LFoundry sensors before and after the increase of the backside implant concentration. 
The dotted lines correspond to the sensors with low backside implant concentration, measured at wafer level. The dashed blue line is from the sensor with increased backside doping concentration, measured at module level after flip-chipping. 
The maximal current allowed for the Phase II requirements is indicated in light grey. All the IV measurements were done at a temperature of 20 $^o$C.}
\end{minipage}
\end{figure}

The first two requirements in Table \ref{Phase2req} constrain the IV curve of the sensors. 
Figure \ref{IV} shows IV measurements of sensors from two different wafers. 
The bare sensors from the first wafer showed a leakage current much higher than the Phase II requirements. This was tracked down to a low implant dose in the sensors backside. Two wafers were re-implanted to a higher dose before post-processing for Under Bump Metallization (UBM).
The IV curve of a sensor with a higher implant concentration after flip-chipping is also shown in Figure \ref{IV}. 
The breakdown voltage and the leakage current at $V_\text{dep}$ + 50 V comply with the CMS Phase II requirements. 

\subsection{Efficiency}
\label{Sec_Eff}

\begin{figure}[H]
\begin{minipage}{.5\textwidth}
\begin{center}
\includegraphics[scale = 0.2]{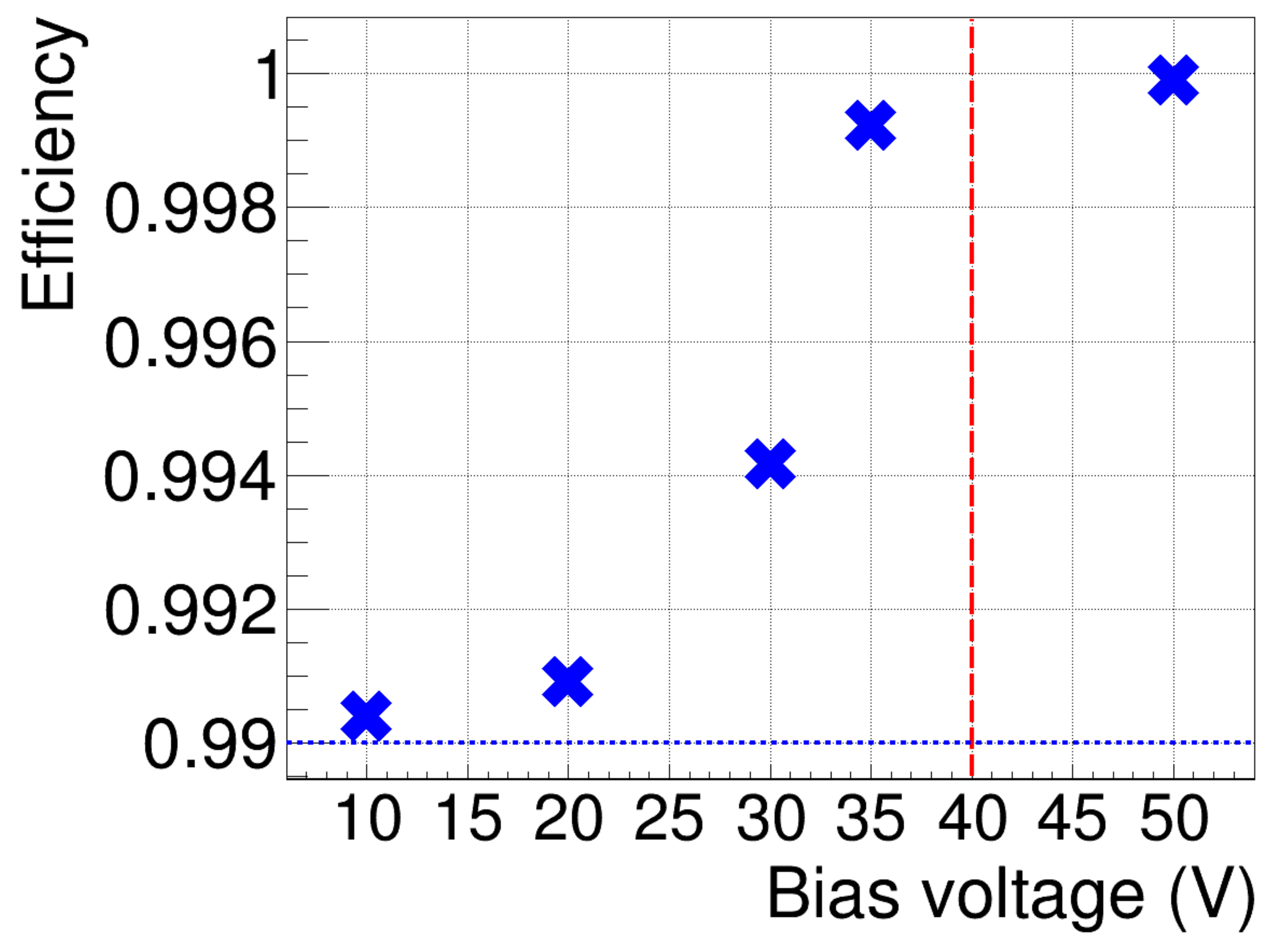}
\end{center}
\caption{\label{EffvsBias}The blue markers show the efficiency of the sensor w.r.t the bias voltage. The blue line shows the 99\% efficiency limit and the red line is positioned at the depletion voltage of the sensor. }
\end{minipage}
\hspace{1.5pc}
\begin{minipage}{.5\textwidth}
  \vspace{-0.5cm}
\begin{center}
\includegraphics[scale = 0.19]{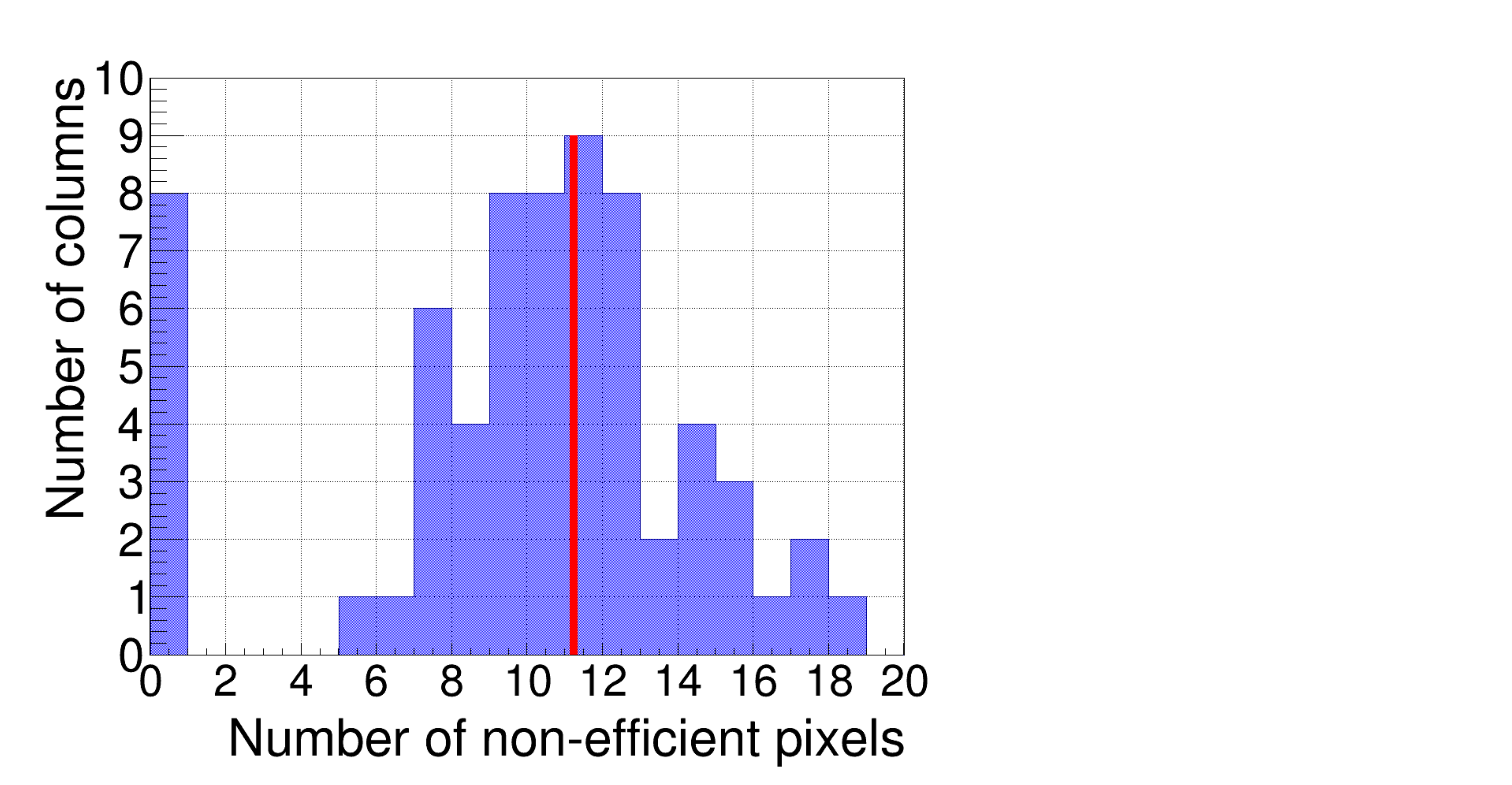}
\end{center}
\caption{\label{StitchedEff} The histogram shows the number of pixels per row that did not detect all telescope hits. The red line shows the same for the two rows adjacent to the stitching line. }
\end{minipage}
\end{figure}

The efficiency is defined as the probability of detecting a hit on the Device Under Test (DUT) within 500 $\upmu$m of each telescope track hit.
The efficiency of a non-irradiated LFoundry passive sensor, interconnected to a RD53A chip, was investigated in a beam test at DESY \cite{DESY} using the EUDET telescope,  with 5.2 GeV electrons. The module was read-out with the BDAQ53 system \cite{BDAQ_gitlab}. 
Figure \ref{EffvsBias} shows that the Phase II efficiency requirement is matched from 10 V. 

The sensor was processed using two reticle blocks stiched together (see Section \ref{CMOSstitching}). To verify that the stitching process does not reduce the efficiency, pixels next to the stitched part of the sensor (stitching line) are compared to the rest of the pixel matrix. 
The number of pixels per row of the sensor that were not fully efficient (under 85\% efficiency for a mean number of hits of 42) is plotted on Figure \ref{StitchedEff}.  The number of not fully efficient pixels per row should be seen as a part of a set of 384 pixels. 
Some rows reach an efficiency of 1 for all pixels but most rows show around 10 not-fully-efficient pixels. The plot also shows this value for the rows adjacent to the stitching line. It is consistent with the values of the non-stitched rows and thus a strong argument for showing that stitching does not noticeably reduce the efficiency.

\subsection{Charge collection}
\label{Sec_Q}

\begin{figure}[H]
\begin{minipage}{.5\textwidth}
\begin{center}
\includegraphics[width = 8cm, height = 5cm]{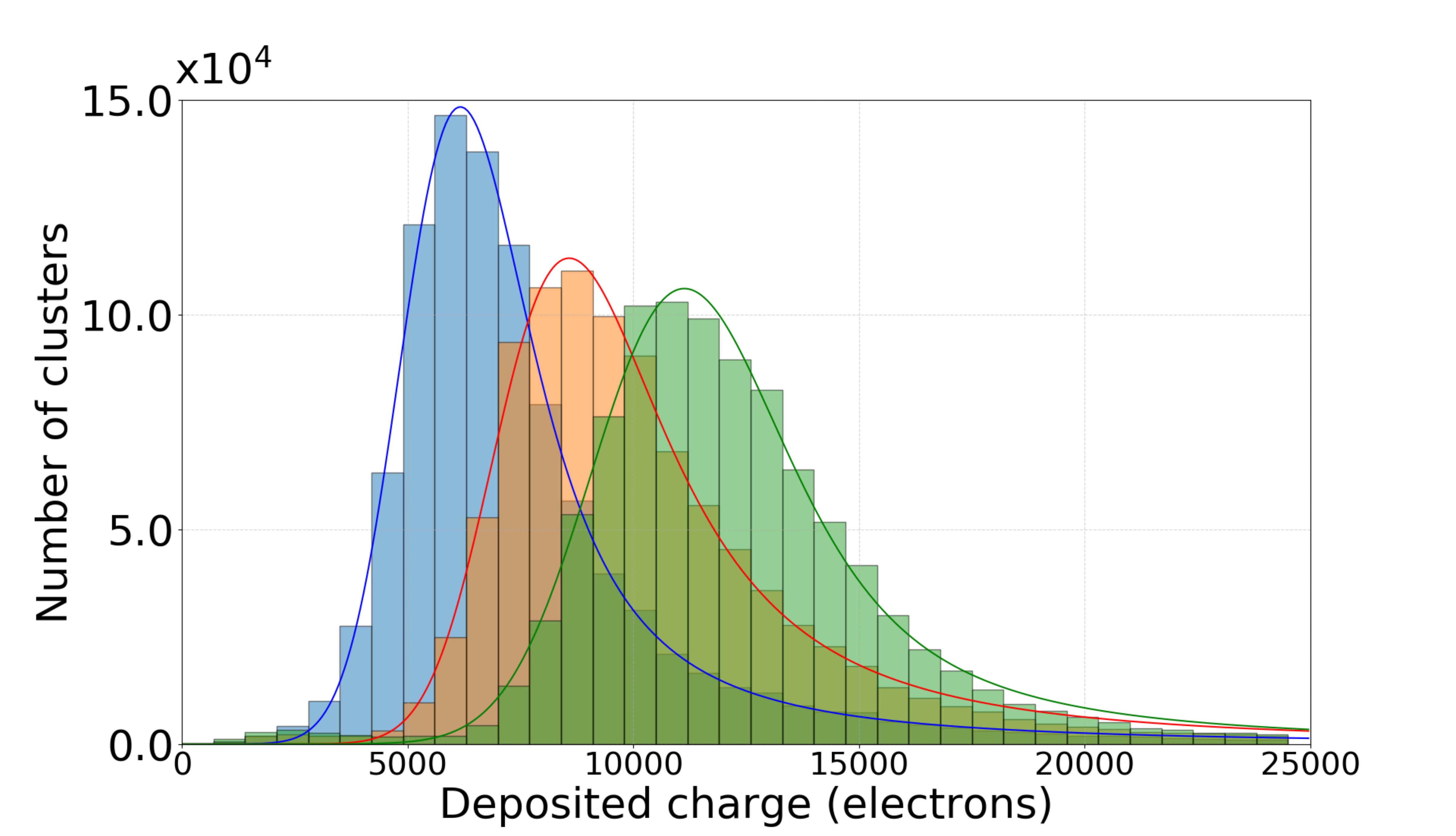}
\end{center}
\caption{\label{LandauElectrons}Cluster charge distribution with respect to the bias voltage. The blue distribution corresponds to a bias of 10 V, the orange one to a bias of 20 V and the green one to a bias of 50 V. The MPVs can be found in Figure \ref{MPVvsBias}.}
\end{minipage}
\hspace{1.5pc}
\begin{minipage}{.5\textwidth}
  \vspace{-0.5cm}
  \begin{center}
\includegraphics[scale = 0.22]{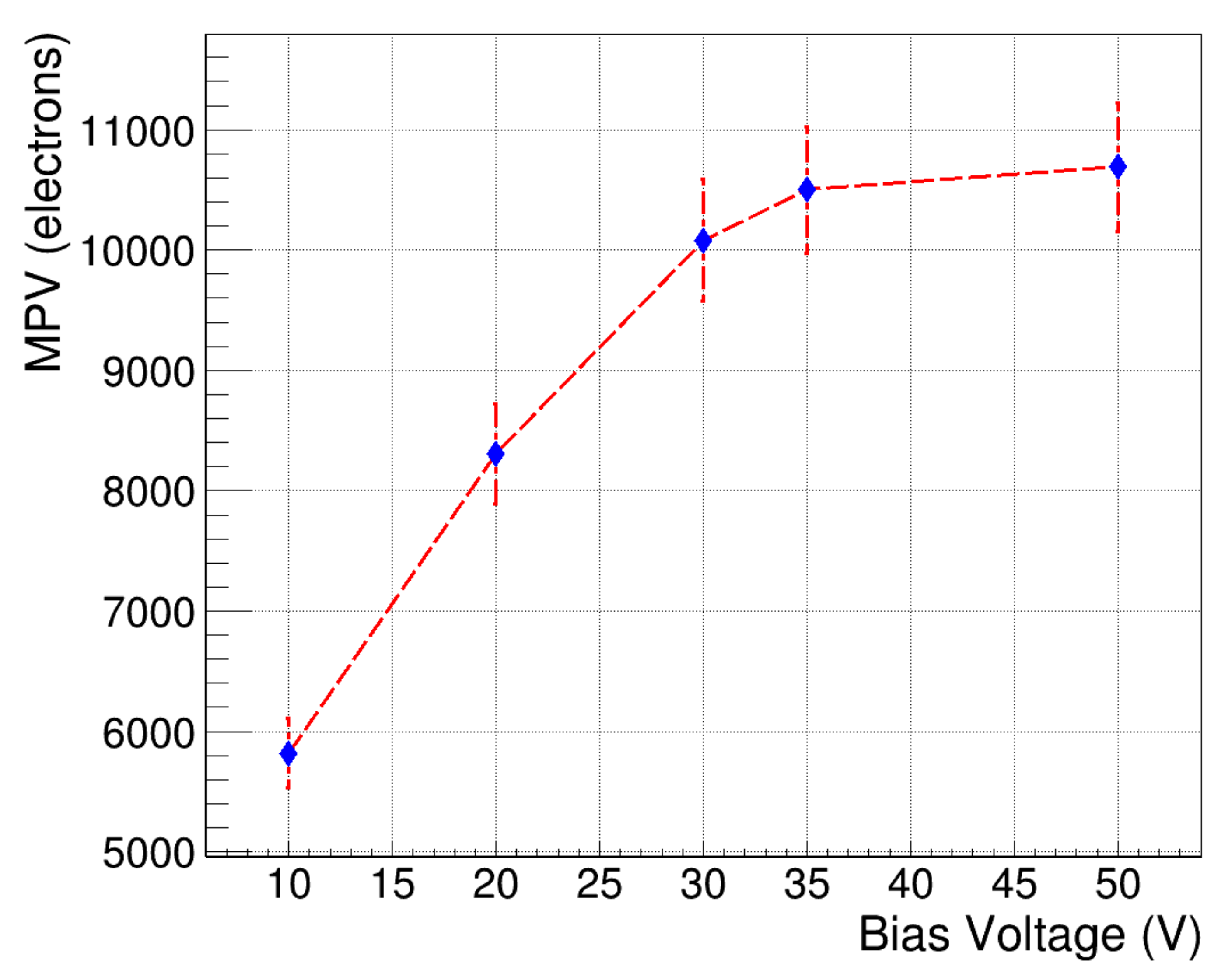}
\end{center}
\caption{\label{MPVvsBias}MPV of the collected charge as a function of the bias voltage after deconvolution of the fitted function. The error bars come from the charge calibration uncertainties.} 
\end{minipage}
\end{figure}

The charge collected from a minimum ionizing particle in a thin silicon sensor is of approximately 70 electrons per $\upmu$m, using Bichsel's formula \cite{VolumeElec}. The expected charge collected in a 150 $\upmu$m thick sensor is thus approximately 10500 electrons.
The distribution of the charge collected in a cluster has been obtained using the Time-over-Threshold (ToT) distribution measured by the RD53A during a testbeam run at perpendicular incidence of the beam on the DUT. The electron beam had an energy of 5.2 GeV. 
The ToT was then converted to a physical charge using an X-ray source. Metal targets (Ag, Cu, Mo, Sn, Zn) were irradiated by this source and emitted transition radiation of known energy towards the RD53A. The deposited charge was then measured and the relation between ToT and charge obtained. 
The most probable deposited charge value (MPV) is obtained by fitting the cluster charge distribution with a convolution of a Gaussian
and a Landau distribution. The fit must then be deconvolved to obtain the MPV of the Landau distribution. Figure \ref{LandauElectrons} shows the convoluted distribution and their fit for different bias voltages. Figure \ref{MPVvsBias} shows the increase of the MPV after deconvolution w.r.t the bias voltage. 

\subsection{Spatial resolution}
\label{Sec_Res}

\begin{figure}[H]
\begin{center}
\includegraphics[scale = 0.2]{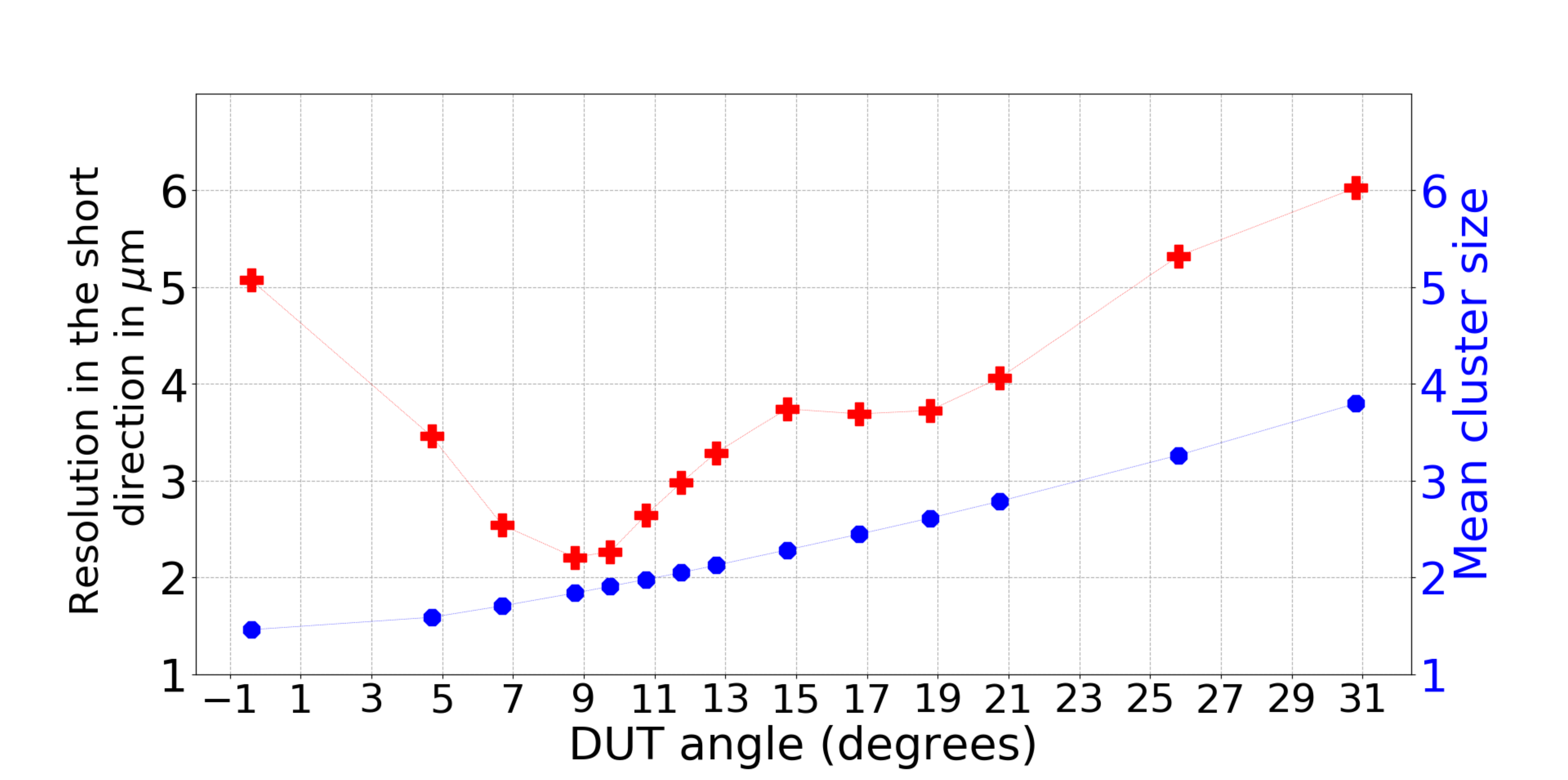}
\end{center}
\caption{\label{DUTres} The red markers show the DUT resolution w.r.t the beam incidence angle in the direction of the short side of the pixel cell. The blue ones show the mean cluster size in the same direction.}
\end{figure}

The hit position on the DUT is independently computed in the two directions, along the short and long pixel cell sides, on the sensor surface.
The following results focus on the resolution along the 25 $\upmu$m side of the sensor. 
The hit position on the DUT is computed using the charge-weighted center of the cluster \cite{COC}. 
The resolution of the telescope has to be taken into account and projected on the sensor plane. Thus, the resolution of the DUT is extracted from the formula 
\begin{equation}
  \sigma_{\Delta x} = \sqrt{\sigma_{\text{DUT}}^2 + \left(  \frac{ \sigma_{\text{telescope}}}{  \cos( \theta )} \right)^2   }
\end{equation}

where $\theta$ is the angle between the beam axis and the sensor and the standard deviation $\sigma_{\Delta x}$ comes from the distribution of the residues between the telescope tracks and the hits on the DUT.
Figure \ref{DUTres} shows the resolution of the sensor as a function of the beam incidence angle as well as the average cluster size along the 25 $\upmu$m side. 
The minimal resolution of 2.2 $\upmu$m is reached between 8.75 and 9.75 degrees. It is compatible with the theoretical value of $\arctan(\frac{p}{T}) \approx 9.5$ degrees which  is the angle maximizing the number of clusters of size 2 in the 25 $\upmu$m direction. 
Here, $p$ denotes the pixel pitch (25 $\upmu$m) and $T$ the thickness of the sensor.

\subsection{Inter-pixel crosstalk}
\label{Sec_X}

\begin{figure}[H]
\begin{minipage}{.5\textwidth}
\begin{center}
\includegraphics[scale = 0.2]{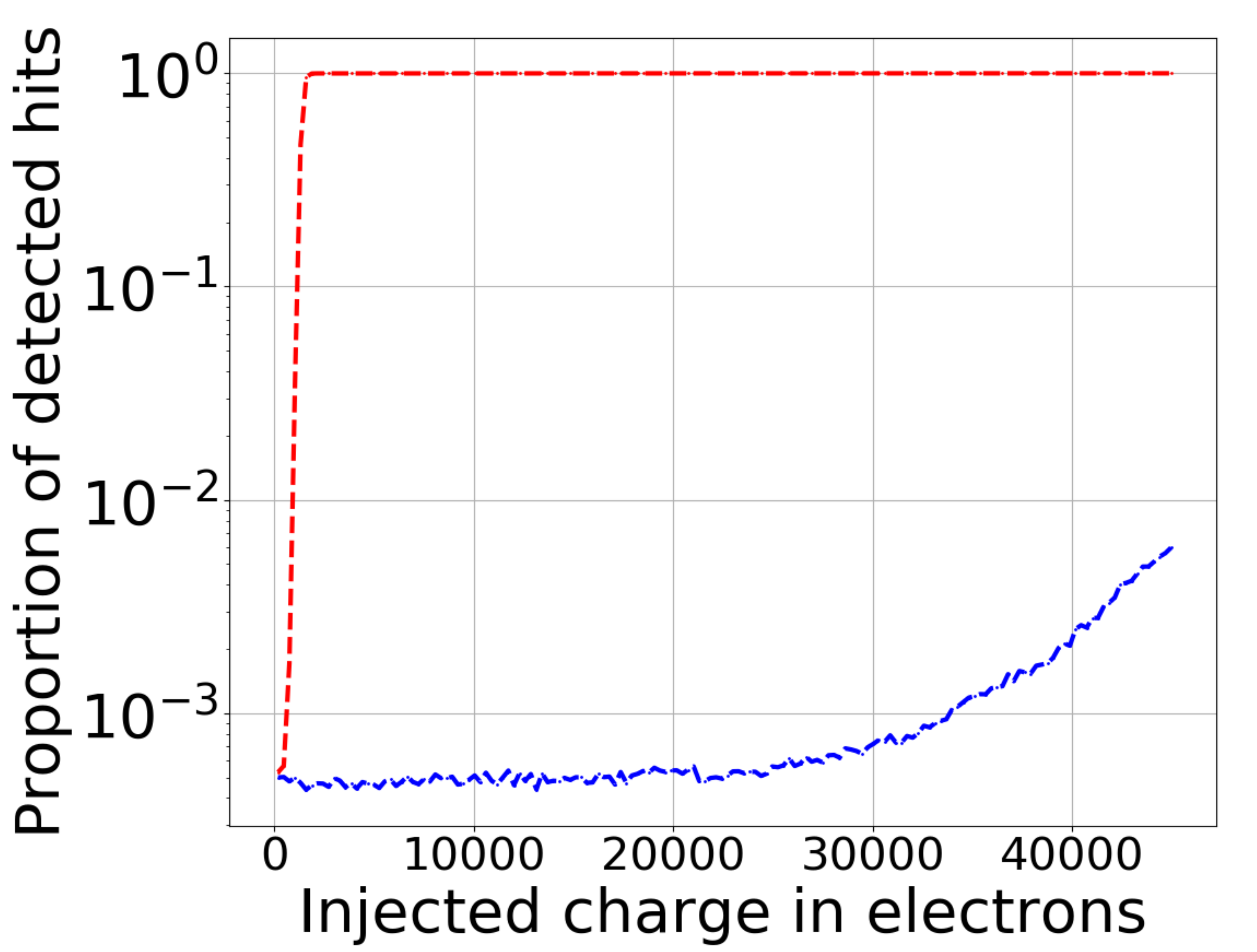}
\end{center}
\caption{\label{ScurveXtalk} Mean proportion of detected hits w.r.t the injected charge. The red curve corresponds to detected hits in the pixels that have been injected with a charge and the blue curve to crosstalk coupled, neighbouring pixels, that have not been injected with a charge.}
\end{minipage}
\hspace{1.5pc}
\begin{minipage}{.5\textwidth}
  \vspace{-1.3cm}
  \begin{center}
\includegraphics[width = 7cm, height = 4cm]{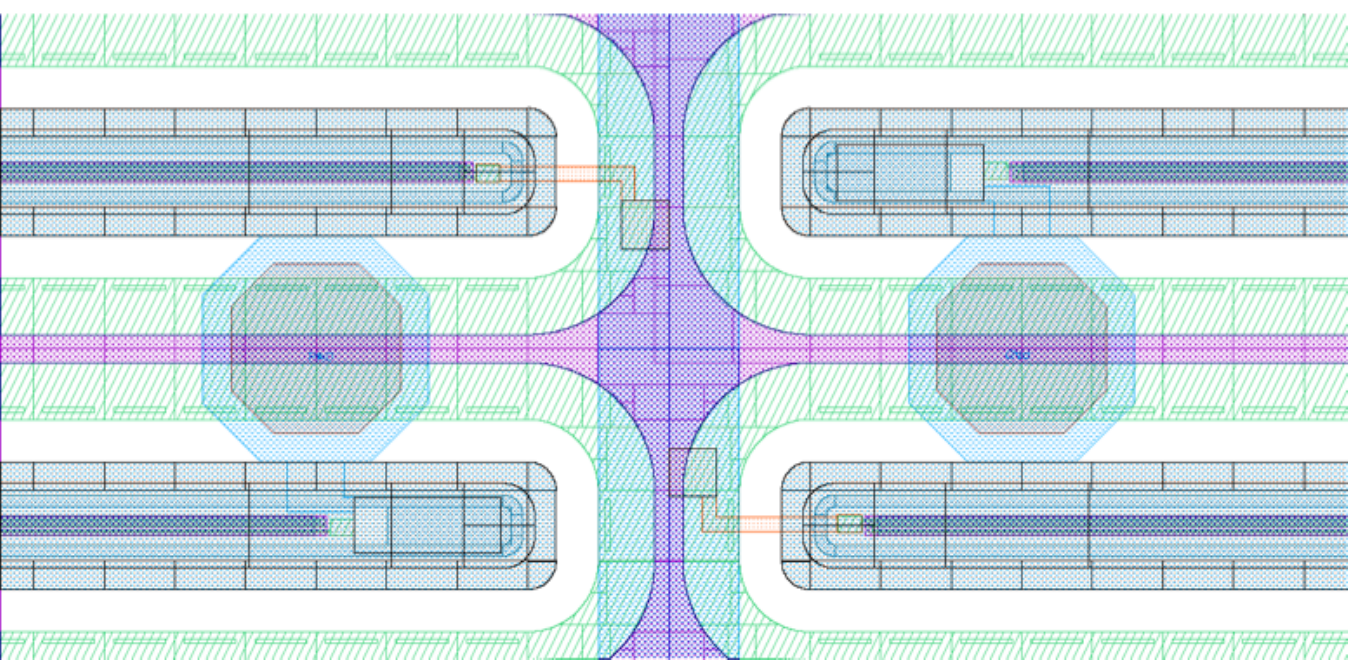}
\end{center}
\caption{\label{BumpBond} LFoundry pixel design showing the proximity of the bump bond pads (octogonal shapes) to the implants (in blue).   }
\end{minipage}
\end{figure}

Capacitive coupling between neighbouring pixels (known as crosstalk) leads to spurious hits and can deteriorate the sensor resolution. 
An enhanced cross-talk has been measured on $25\times100$ $\upmu$m$^2$ sensors of other producers with respect to the $50\times50$ $\upmu$m$^2$ geometry \cite{CMSplanar}. This was due to the overlay or close vicinity of the metal system of the UBM of one pixel with the implant layer of the neighbouring pixel, see Figure \ref{BumpBond}.
Crosstalk can be computed by injecting a known charge through the RD53A and reading out the neighbouring pixels. 
Figure \ref{ScurveXtalk} shows the proportion of detected hits in pixels that have received a charge injection as well as those that did not but are crosstalk-coupled to the injected ones. The former detect all hits after their threshold charge $T_1$ is crossed 
while the latter do not reach the 50 \% detection efficiency point (which defines the second threshold $T_2$) even at the highest charge that can be injected using the RD53A. 
The crosstalk $\alpha$ is defined as  

\begin{equation}
  \alpha  = \frac{1}{1+ \frac{T_2}{T_1}}  < 2.5  \%  
 \end{equation}

and is calculated by inserting the values of the measured threshold $T_1$ and the highest charge $T_2$ injected through the ROC.

\section{Conclusion}

The first laboratory and TB measurements of an unirradiated LFoundry passive CMOS sensor flip-chipped to a RD53A ROC show promising results towards the application in the Phase II upgrade of the CMS IT. 
No noticeable loss of efficiency is found due to the stitching process. The crosstalk is shown to be smaller than 2.5\% and the minimal resolution along the short side of the pixel cell of the sensor reaches 2.2 $\upmu$m.
RD53A modules assembled with LFoundry sensors will be characterized after irradiation to the highest fluence level foreseen in the CMS IT system.

\section{References}

\bibliographystyle{iopart-num}
\bibliography{Paper}

\ack

The measurements leading to these results have been performed at the Test Beam Facility at DESY Hamburg (Germany), a member of the Helmholtz Association (HGF).

\end{document}